# Improving Tag-Clouds as Visual Information Retrieval Interfaces

Yusef Hassan-Montero[a1,] and Víctor Herrero-Solana[a]

[a] *Scimago Research Group. University of Granada, Faculty of Library and Information Science, Colegio Máximo de Cartuja s.n., 18071 Granada, SPAIN*

Tagging-based systems enable users to categorize web resources by means of tags (freely chosen keywords), in order to re-finding these resources later. Tagging is implicitly also a social indexing process, since users share their tags and resources, constructing a social tag index, so-called folksonomy. At the same time of tagging-based system, has been popularised an interface model for visual information retrieval known as Tag-Cloud. In this model, the most frequently used tags are displayed in alphabetical order. This paper presents a novel approach to Tag-Cloud's tags selection, and proposes the use of clustering algorithms for visual layout, with the aim of improve browsing experience. The results suggest that presented approach reduces the semantic density of tag set, and improves the visual consistency of Tag-Cloud layout.

## 1 INTRODUCTION

With the advent of social bookmarking tools, such as del.icio.us or connotea [9], a new approach for metadata creation has emerged, known as Tagging. Tagging-based systems enable users to add tags – freely chosen keywords – to web resources for categorize these resources in order to re-find them later.

Tagging is not only an individual process of categorization, but implicitly it is also a social process of indexing, a social process of knowledge construction. Users share their resources with their tags, generating an aggregated tag-index so-called folksonomy. Folksonomy term, coined by Thomas Vander Wal in AIfIA mailing list, is a one-word neologism that comes from the words 'taxonomy' and 'folk' [15]. There exists some discussion about its accuracy, although we prefer to use it due to its popularity and widely acceptance.

Folksonomy allows anyone to access to any web resource that was previously tagged, based on two main paradigms of information access: Information Filtering (IF) and Information Retrieval (IR).

In IF user plays a passive role, expecting that system *pushes* or sends toward him information of interest according to some previously defined profile. Social bookmarking tools allow a simple IF access model, where user can subscribe to a set of specific tags via RSS/Atom syndication, and thus be alerted when a new resource will be indexed with this set.

On the other hand, in IR user seeks actively information, *pulling* at it, by means of querying or browsing. In tag querying, user enters one or more tags in the search box to obtain an ordered list of resources which were in relation with these tags. When a user is scanning this list, the system also provide a list of related tags (i.e. tags with a high degree of co-occurrence with the original tag), allowing hypertext browsing.

In order to enable visual browsing, social bookmarking tools typically provide an interface model known as Tag-Cloud (figure 1). A Tag-Cloud is a list of the most popular tags, usually displayed in alphabetical order, and visually weighted by font size. In a Tag-Cloud, when a user clicks on tag obtains an ordered list of tag-described resources, as well as a list of others related tags. Whereas querying requires to user to formulate previously his information needs, visual browsing allows the user to recognize his information needs scanning the interface. Visual browsing is similar to hypertext browsing in the way that both of which allow the user to search by browsing. However there is a difference, that is, visual interfaces provide a global view of tags or resources collection, a contextual view.

Tag-Cloud is a simple and widely used visual interface model, but with some restrictions that limit its utility as visual information retrieval interface. It is due to:

- The method to select the tag set to display is based exclusively on the use frequency, which inevitably entails that displayed tags have a high semantic density. In terms of discrimination value, the most frequently-used terms are the worst discriminators [16]. As indicate Begelman, Keller and Smadja [1], very few different topics, with all their related tags, tend to dominate the whole cloud. Xu et al. [26] suggest also the need of research on tag selection methods in order to improve Tag-Clouds.
- Alphabetical arrangements of displayed tags neither facilitate visual scanning nor enable infer semantic relation between tags. We defend that similarity-based layout would improve Tag-Cloud browsing. Although a folksonomy is commonly defined as a flat space of keywords without previously defined semantic relationships between tags, different studies [12] [3] [20] demonstrate that associative and hierarchical relationships of similarity between tags can be inferred from tag co-occurrence analysis [17].

---
[1] Corresponding Author: Yusef Hassan-Montero. Email: yusefhassan@gmail.com

Figure 1: Traditional Tag-Cloud. Tags have been selected and visually weighted according to its frequency of use.

In this paper we propose an alternative method for tag selection and display. Our method reduces the semantic density of the tag set (overlapping between tags). Also, we propose an alternative method for tags layout, where tags are grouped by similarity, based on clustering techniques and co-occurrence analysis.

## 2 BACKGROUND

Polysemy, synonym and ego-oriented nature of many tags (e.g. 'toread', 'me', 'todo'), are well known problems of tagging in order to support IR. Furthermore, a large volume of tags identify resource's author (e.g. 'nielsen'), subjective opinion of tagger (e.g. 'cool'), or kind of resource (e.g. 'book', 'blog') [7][26][3]. In other words, tags are not (always) topic or subject index terms.

Guy and Tonkin [8] discuss how to improve tags by means of tagging literacy, concluding with an interesting statement: "Possibly the real problem with folksonomies is not their chaotic tags but that they are trying to serve two masters at once; the personal collection, and the collective collection. Is it possible to have the best of both worlds?"

Wu, Zhang and Yu [25] demonstrate that is possible to extract tags with collective usefulness from the sum of individual and freely assigned tags, solving automatically tag ambiguity problems. Furthermore, from our point of view, tagging systems enclose important and exclusive strengths, some of which result from the chaotic, uncontrolled and free nature of tagging:

- Folksonomies directly reflects the vocabulary of users, enabling matching users' real needs and language [10][15]. The best way to obtain a user-centered indexing [6], is through user-generated indexing.
- In addition, given that folksonomy emerge from collective agreement, tags have more accurate, truthful and democratic meaning than the assigned by a single person (author, professional indexer, etc.). As say Wu, Zhang and Yu [25], users negotiate the meaning of tags in an implicit asymmetric communication. According to "use" theory of meaning proposed by Ludwig Wittgenstein philosopher, meaning of a word is in its use. Meaning, as a social event, happens among language users.
- An inherent problem of human indexing process is the called "inter-indexer inconsistency". It happens when different indexers use different index terms to describe the same document [14]. This problem would be reduced when indexing is obtained by aggregation, such as occur in social tagging. As reveals Golder and Huberman [7] in their study, usually after 100 or so users index the same resource, each tag's frequency is a nearly fixed proportion of the total frequency of all tags assigned to the resource. Since tag proportions tend to stabilize, it means that a unique tag weight might be assigned to each resource.
- Folksonomies allow information discovery by serendipity [10][15].
- In the greater part of cases, where tagging could be used, other solutions such as controlled vocabulary are impossible to implement. In words of Quintarelli [15], "folksonomies are better than nothing".

Despite the potential of tagging systems for IR, there's not enough research about the effectiveness and usefulness of folksonomies.

Tagging effectiveness can be measured by means of two related parameters: term/tag specificity and indexing/tagging exhaustivity. These two variables indicate the number of resources described by one tag, and the number of tags assigned to one resource respectively [22]. Through a broad tag, user will retrieve many relevant resources, but an important number of irrelevant ones, at the same time. Narrow tags retrieve few and mainly relevant resources, but may miss some relevant resources. In terms of traditional IR, broader tags entail high recall and low precision, whereas narrower tags entail low recall and high precision.

Xu et al. [26] relate broader tags to discovery tasks (i.e. find new resources), and narrow tags to recovery tasks (i.e. find resources previously discovered). Likewise, it's reasonable to connect broader tags to more general purpose searching tasks, and narrow tags to more specific and goal-oriented searching tasks. In other words, browsing and querying respectively.

Therefore, the main question is: do folksonomies' tags tend to be broader or narrower? And thus, are folksonomies more suitable to support browsing or querying?

Despite the relation between number of resources and the volume tags used to describe them by an user are not well correlated - i.e. some users use large tag sets and others small tag sets- [7], Brooks and Montanez [3] reveal that taggers tend to assign broader tags. Tagging, also, is a low exhaustive indexing process, since at 90% of cases the average user assigns less than 5 tags per resource [24]. These facts are reasonable if we consider that the low cognitive cost of tagging is one of the main factors of its popularity [21]. In this sense, to assign few, broader and inclusive tags would demand less cognitive effort than other kind of tags, narrower and more exclusive. It therefore follows that the low-specific nature of tags makes them more suitable to support browsing than querying.

There are some interesting works about folksonomies visualization, such as the work of Dubinko et al. [5] where authors present a novel approach to visualize the evolution of tags over the time. However, a few exceptions are destined to IR.

A case in question is the tag browser proposed by Shaw [19], a Tag-Cloud mapped like a graph, where tags are represented as visually distributed nodes, and similarity relationships as edges between nodes. On the other hand, Bielenberg and Zacher [2] present a Tag-Cloud with circular form, where font size and distance to the centre represent the importance of a tag, but where distance between tags doesn't represent their similarity.

## 3 MATERIALS AND METHODS

### 3.1 Sample

Our study was performed on a large sample downloaded from 'del.icio.us' bookmarking tool, during October 2005. This sample contains 218,063 URLs tagged with 242,349 tags by 111,234 users, and was retrieved by means of ad-hoc programmed crawler.

Del.icio.us' tagging system is characterized by allowing free-for-all tagging - i.e. any user can tag any web resource- and blind tagging - i.e. user cannot view other tags assigned to the same web resource while tagging- [11]. For these and other reasons [4], del.icio.us probably is the best source of data for tagging research.

### 3.2 Tag Similarity

Tag similarity is the main concept of our approach, thus we first and foremost will define it.

Similarity between two tags can be measured by different forms. The easiest method is counting the number of co-occurrences, that is, the number of times when two tags are assigned to the same resource. Such as Cattuto, Loreto and Pietronero [4] show, the non-trivial nature of co-occurrence relationships between tags might be ascribed to semantics.

In this paper, tag similarity is considered a kind of semantic relationship between tags, measured by means of relative co-occurrence between tags, known as Jaccard coefficient. Let A and B be the sets of resources described by two tags, relative co-occurrence is defined as:

$$RC(A,B) = \frac{|A \cap B|}{|A \cup B|} \quad (1)$$

That is, relative co-occurrence is equal to the division between the number of resources in which tags co-occur, and the number of resources in which appear any one of two tags. In this paper, similarity based on co-occurrence is also called tag overlapping.

### 3.3. Tag Selection

Commonly, Tag-Clouds' tags are selected on the basis of its frequency of use, without considering others complementary variables. This selection method entails that Tag-Clouds offers a semantically homogeneous image, where almost all tags are similar to each other.

Despite term weighting and selecting have been studied in great depth in IR field, generally these solutions are focus to selecting terms from a single document [13]. However, our approach aims to select the tags that better characterize the whole collection of tagged resources.

For this purpose, tag usefulness is determined by: (1) its capacity to represent each resource as compared to other tags assigned to the same resource; (2) the volume of covered resources as compared to other tags; (2) its capacity to cover these resources less covered by other tags.

If we consider folksonomy as a vector space of resources $D_i = (d_{i0}, \ldots, d_{in})$, each one characterized trough one or more tags $T_j = (t_{j0}, \ldots, t_{jm})$ weighting according to the number of tag's users, then $d_{ij}$ represents the frequency with which $T_j$ tag has been used to describe $D_i$ resource. Let $F(T_j)$ denote the usefulness of $T_j$ tag to be part of Tag-Cloud, as follows:

$$F(T_j) = \sum_{i=1}^{i=n} \left( \frac{\log(d_{ij})}{m^2} \right) \quad (2)$$

Where $n$ is the number of resources which have been described by $T_j$ tag, and $m$ is the number of different tags assigned to $D_i$ resource.

This function is a summation of a simplified version of length-normalized *tf·idf* function [18], but where $\log(d_{ij})$ diminishes the effect of high frequently used tags within a resource, and the square of $m$ diminishes the effect of low discriminative tags. Whereas in length-normalized *tf·idf* function the denominator was used to ensure that all documents have equal chance of being retrieved, in this present function is used to give better weighting to these tags which describe resources which are worse covered by other tags. In other words, numerator measures the representation value of tag, while denominator measures its potential discrimination value.

Table 1. Selection method comparison over 95 highest weighted tags by each function. Coverage is the number of resources that have been described by at least one of the 95 selected tags. Overlapping is the relative co-occurrence between tags.

|  | Coverage | Overlapping average | Overlapping standard deviation |
|---|---|---|---|
| a) $F(T_j) = n$ | 188,761 (86.56%) | 0.0503 | 0.0414 |
| b) $F(T_j) = \sum_{i=1}^{i=n} (d_{ij})$ | 187,907 (86.17%) | 0.0399 | 0.0425 |
| c) $F(T_j) = \sum_{i=1}^{i=n} \left( \frac{\log(d_{ij})}{m} \right)$ | 191,567 (87.85%) | 0.0329 | 0.0406 |
| d) $F(T_j) = \sum_{i=1}^{i=n} \left( \frac{\log(d_{ij})}{m^2} \right)$ | 190,405 (87.32%) | 0.0242 | 0.0372 |

As it is shown in table 1, in terms of overlapping between tags, our approach (d) offers better results than traditional selection methods (a, b) and than the same function without square of *m* (c). Differences between functions' coverage are minimal, due to the powerful coverage of the most frequently used tags (e.g. 'web', 'design', 'programming'…) which are present within all selections.

### 3.4. Tag-Cloud Layout

Alphabetical-based schemes are useful for know-item searching, i.e. when user knows previously what tag he is looking for, such as when user browses his personal Tag-Cloud. However, in the case of a user browsing a collectively generated and previously unknown Tag-Cloud, it would be perceived as a chaotic tag soup.

Our proposal for Tag-Cloud layout is based on the assumption that clustering techniques can improve Tag-Clouds' browsing experience [1][2]. Data clustering comprise a collection of unsupervised learning algorithms which, by means of a iterative process, aim to group a set of objects into clusters, whose members are similar to each other and dissimilar to members of others clusters.

On the *N*-dimensional similarity matrix of tags - computed with the relative co-occurrence function (1)-, where *N* is the number of different tags, we can already apply a clustering algorithm in order to classify tags in clusters. For this purpose we have chosen bisecting K-means algorithm - a divisive (non-agglomerative) hierarchical clustering algorithm explained in detail by Steinbach, Karypis and Kumar [23] - using cosine similarity function to measure co-occurrence patterns between tags. In present experiment *N*=95 and *K*=12, where *N* is the number of highest weighted tags to cluster, and *K* is the predefined number of clusters to obtain.

The display method is similar to traditional Tag-Cloud layout, with the difference that tags are grouped with semantically similar tags, and likewise clusters of tags are displayed near semantically similar clusters. Similar tags are horizontally neighbours, whereas similar clusters are vertically neighbours.

## 4 RESULTS

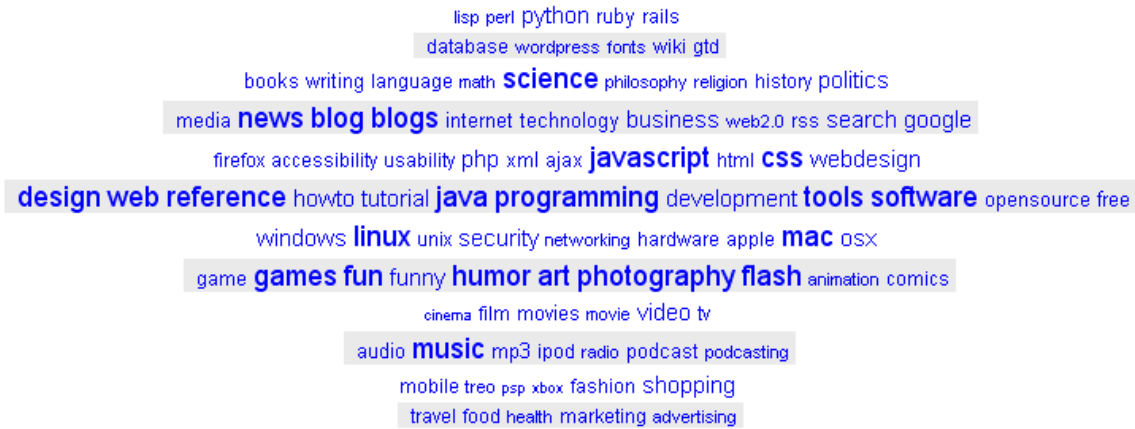

Figure 2: Improved Tag-Cloud. Tags have been selected and visually weighted according to function 1.

The improved Tag-Cloud (figure 2) share 65 tags with the traditional Tag-Cloud (figure 1). From de traditional Tag-Cloud has disappeared the less significant tags (e.g. 'toread', 'diy' or 'cool'), some specific names of sites (e.g. 'del.icio.us', 'delicious' or 'flickr'), and several synonymous tags of the preserved tags (e.g. 'webdev', 'photos', 'blogging', 'tech' or 'tutorials').

On the other hand, in the improved Tag-Cloud numerous tags with higher meaning have emerged. Some examples are: 'philosophy', 'religion', 'health', 'food', 'fashion', 'math', 'accessibility', 'radio', 'cinema', etc. These new tags have more discrimination value and increase the presence of topics not related to web design and technologies, reducing semantic density. Nevertheless, tech-topics still dominate the Tag-Cloud, due to the common interest of del.icio.us' tech-savvy user base [10].

Clustering offers more coherent visual distribution of tags than traditional alphabetical arrangements, allowing to differentiate among main topics in Tag-Cloud, as well as to infer semantic knowledge from the neighbours' relationships. For example, if a user doesn't know what 'ajax' is, browsing the interface he might suppose that it is something conceptually related to 'javascript' and 'xml'.

## 5 DISCUSSION

In this paper we have presented a simple and low computational cost method to select the *N* most relevant tags from a collaborative tag index, and a technique to display these tags with clustering-based layout.

Despite the improvements of our proposal, synonym and plural forms problems still remain, but we believe that it is a problem with a relative easy solution. Opposite of automatic text indexing, where synonymous terms don't tend to co-occur in the same document, in social tagging systems, since indexing is performed by multiple users, synonymous tags frequently co-occur in the same resource. The simplest solution to this issue could be a predefined similarity threshold which would determines when two tags can be considered synonymous. As it can be seen in figure 2, synonymous or very semantically similar tags appear together as result of its high co-occurrence frequency.

Although in this paper we have only discussed how to improve main-Tag-Clouds, this approach could be used to offer hierarchical navigation by means of sub-Tag-Clouds. For example, when user does click on a tag, he could access to a new Tag-Cloud with the N most relevant tags assigned to these resources described by the clicked tag (without considering it).

Our aim in future works is to improve tag selection and weighting methods, and to propose alternative approaches and graphic metaphors to visualize large folksonomies spaces.